\documentstyle[a4p,doublespace,12pt]{article}
\newcommand{\gtrsim}{\;\raisebox{-0.9ex}
           {$\textstyle\stackrel{\textstyle >}{\sim}$}\;}

\newcommand{\rhalos}{\rho_{{\rm h}\odot}}

\newcommand{\pbar}{\mbox{$\bar{p}$}}

\newcommand{\mzero}{m_{0}}
\newcommand{\mhalf}{m_{1/2}}
\newcommand{\mchi}{m_{\chi}}
\newcommand{\mlcsp}{m_{\rm LCSP}}
\newcommand{\mtop}{m_{t}}
\newcommand{\tbeta}{\tan\beta}
\newcommand{\Ochi}{\Omega_{\chi}}
\newcommand{\sigv}{\langle\sigma v\rangle_{\rm h}}
\newcommand{\tautau}{\tau^{+}\tau^{-}}
\newcommand{\bbbar}{b\bar{b}}

\newcommand{\charg}{\chi_{1}^{\pm}}
\newcommand{\stau}{\tilde{\tau}_{1}}
\newcommand{\Bbbbar}{B_{\bbbar}}

\begin{document}
\begin{titlepage}
\begin{tabbing}
\` UT-ICEPP 96-03\\
\` August 1996\\
\end{tabbing}
%
\begin{center}{\large\bf
Expected Enhancement of the Primary Antiproton Flux 
at the Solar Minimum
}\end{center}
\bigskip
\begin{center}{\large 
T. Mitsui \footnote{E-mail address: mitsui@icepp.s.u-tokyo.ac.jp}, 
K. Maki \footnote{E-mail address: maki@icepp.s.u-tokyo.ac.jp},
and S. Orito \footnote{E-mail address: oriton@icepp.s.u-tokyo.ac.jp}
}\end{center}
\begin{center}{\it  
Department of Physics, School of Science, 
University of Tokyo, Tokyo 113, Japan}\end{center}
\bigskip
\bigskip
\begin{abstract}
We calculate the solar-modulated energy spectra of cosmic-ray
antiprotons ($\bar{p}$'s) from two candidate primary
sources, i.e., evaporating primordial black holes and the annihilation
of neutralino dark matter, as well as for the secondary
$\bar{p}$'s produced by cosmic-ray interactions with interstellar gas.
A large enhancement toward the solar minimum phase emerges in the
low-energy flux of $\bar{p}$'s from the primary sources,
whereas the flux of the secondary $\bar{p}$'s,
falling steeply below 2 GeV,
does not significantly vary.
This
enables us to conduct a very sensitive search for
primary $\bar{p}$ components by
precisely measuring the $\bar{p}$ spectrum, especially
at low energies,
throughout the forthcoming solar minimum phase.
\end{abstract}
\end{titlepage}

\section{Introduction}

Since the discovery of antiprotons (\pbar's) in cosmic rays 
\cite{G84,obs,BESS93}, their origin has attracted much interest 
(for a review, see Ref.~\cite{ste_pb}).  From a standpoint of cosmology 
and elementary particle physics, it would be of great importance to detect 
or to set a stringent limit on the primary \pbar ~components from such 
sources as the annihilation of neutralino dark matter 
\cite{SS84,JK94,BFFM95,DKKW95}, evaporating primordial black holes (PBHs) 
\cite{KSWW,MMO95} and superconducting cosmic strings \cite{SCS}.  No 
evidence for such primary \pbar ~components has been obtained so far.

For the very sensitive search for primary \pbar's anticipated in the near
future experiments \cite{BESSlong}, a potential source of background could 
be the secondary \pbar's produced by cosmic-ray interactions with 
interstellar gas, since the models of the cosmic-ray propagation predict 
the secondary \pbar ~flux at a level close to the present observations 
\cite{ste_pb,webb_pb,gaiss,simon_lb,Mitsui}.  However, the energy spectra 
of the primary \pbar's are generally expected to be rather soft (i.e., flux 
increases with decreasing energy), whereas the spectra of the secondary 
\pbar's should steeply fall below 2 GeV due to the kinematics of the 
\pbar ~production.  The signals of the primary \pbar's can, therefore, be 
searched for in the low-energy regions.

In this Letter, we point out that such low-energy portion of the primary 
\pbar's should show a large enhancement toward the solar minimum phase and 
a rapid decrease afterward, whereas the secondary component, falling steeply 
below 2 GeV, does not significantly vary.  The next solar minimum is 
expected to arrive in a few years: From the observations of 22-yr solar 
cycle over the past 45 years by, e.g., the CLIMAX neutron monitor 
\cite{climax}, one can expect that the next solar minimum should be arriving 
in the years 1997 to 1998, which should be followed by a rapid increase of 
solar activity.  The low-energy flux of the primary \pbar's, e.g., those 
from evaporating PBHs, would increase by a factor of 5 in 1997--1998 as 
compared to the level in 1993, and would then rapidly decrease by a factor 
of 15 in 2000.  On the other hand, the secondary \pbar ~component shall 
remain stable within a factor of 1.5 throughout this period.  Therefore the 
extraction of the primary \pbar ~components would become possible with 
precision measurements of the \pbar ~spectra over the next several years, 
providing a unique opportunity either to detect or put a stringent limit on 
the primary \pbar ~components.

\section{Evaporating primordial black holes}

Primordial black holes (PBHs) may exist as a result of initial density 
fluctuations, phase transitions, or the collapse of cosmic strings in the 
early Universe (for a review, see Ref.~\cite{C85}).  Black holes emit 
particles and evaporate by quantum effects (Hawking radiation) \cite{H74}; 
in particular, PBHs are the only ones with a mass small enough for the 
quantum emission rate to be significant, possibly producing an observable 
flux of $\pbar$'s \cite{KSWW,MMO95}.  

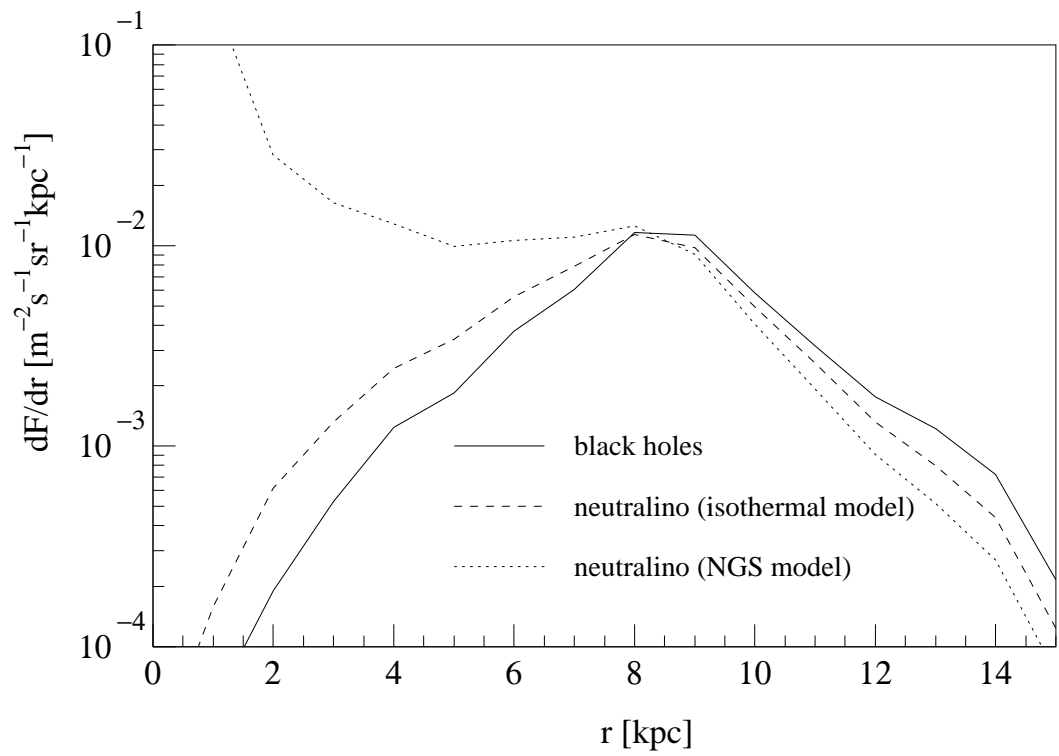
\begin{figure}[h]
\caption{
\protect\baselineskip 10mm
Distribution (along the Galactocentric distance $r$) of the novel primary 
sources (the neutralino annihilation and evaporating PBHs) contributing 
to the integrated $\bar{p}$ flux $F$ at the Solar system, which is 
located at $r=8.5$ kpc.}
\label{fig1}
\end{figure}

In Ref.~\cite{MMO95}, we investigated low-energy cosmic-ray $\bar{p}$'s 
from evaporating PBHs using the Monte Carlo simulation code JETSET 7.4 
\cite{LUND} to obtain the source spectrum of $\pbar$'s from PBHs.  The local 
interstellar flux of these $\bar{p}$'s was then calculated using a 3-D Monte 
Carlo simulation based on the diffusion model of the cosmic-ray propagation.  
The spatial distribution of PBHs was assumed to be proportional to the 
isothermal distribution of halo dark mass, i.e., 
$\rho_{\rm h}(r)=\rhalos(r_{\odot}^{2}+r_{\rm c}^{2})/(r^{2}+r_{\rm c}^{2})$ 
\cite{CO81}, where $r$ is the Galactocentric distance with $r_{\odot}=8.5$ 
kpc being the position of the Solar system, $r_{\rm c}=7.8$ kpc is the core 
radius, and $\rhalos=0.3$ GeV~cm$^{-3}$ is the local halo density.  Our 3-D 
simulation showed that only PBHs within a few kpc of the Solar system 
contribute substantially to the local interstellar $\bar{p}$ flux (see also 
Fig.~\ref{fig1}).  We then found that the local PBH explosion rate, 
${\cal R}$, must be less than $1.7\times10^{-2}$ pc$^{-3}$yr$^{-1}$ in order 
not to conflict with the BESS '93 data \cite{BESS93}.  Nevertheless, we will 
show later that future experiments can have sensitivity to ${\cal R}$ value 
down to $\sim1/30$ of the above upper limit.  In this Letter, we use the 
local interstellar $\bar{p}$ flux obtained in Ref.~\cite{MMO95} after scaling 
it for an appropriate value of ${\cal R}$.   

\section{Annihilation of neutralino dark matter}

Supersymmetric (SUSY) models for particle physics have been extensively 
studied in recent years \cite{SUSY}.  In most models, the lightest 
neutralino $\chi$ is taken to be stable and thus serve as cold dark matter 
(CDM) in the Universe.  If the neutralinos comprise the Galactic halo, their 
annihilation may produce a detectable flux of $\pbar$'s 
\cite{SS84,JK94,BFFM95,DKKW95}, since a pair of neutralinos annihilates into 
quarks, leptons, gauge bosons or Higgs bosons, which then decay or fragment 
into various particles including $\pbar$'s. 

Here we study this process in the framework of minimal $N=1$ supergravity 
models with radiative breaking of the electroweak gauge symmetry 
\cite{SUGRA}.  In this framework, one can calculate all SUSY particle masses 
and couplings by solving the renormalization group equations with only four 
SUSY soft breaking parameters: $\mzero$, $\mhalf$, $\tbeta$ and $A_{0}$.  
The higgsino mass parameter $\mu$ is also determined, except for its sign, 
from the requirement of the electroweak gauge symmetry breaking.  We assume 
$\mu$ to be positive, and the top quark mass to be $\mtop=175$ GeV.  Part of 
the parameter space has already been excluded by theoretical and experimental 
bounds, including recent results of searches for SUSY particles at Tevatron 
\cite{sqTEV,stopTEV} and LEP 1.5 \cite{LEP1.5}.  

We then compute the value of $\Omega_{\chi}h^{2}$, where $\Omega_{\chi}$ is 
the neutralino relic density in units of the critical density of the Universe 
and $h$ is the Hubble constant in units of 100 km~s$^{-1}$Mpc$^{-1}$.  
Recently, Berezinsky et al.\ \cite{B96} have pointed out that the expected 
value of $\Omega_{\chi}h^{2}$ falls within the range of $0.2\pm0.1$ in most 
cosmological models from the viewpoint of the age of the Universe.  Thus, we 
choose here four representative parameter sets which give 
$\Omega_{\chi}h^{2}=0.18$ (see Table~\ref{tab1}).  Neutralino dark matter 
with this value of $\Omega_{\chi}h^{2}$ is most likely pure bino, and 
annihilates predominantly into bottom quarks ($b\bar{b}$) or tau leptons 
($\tautau$).  

To obtain the source spectrum of $\pbar$'s from the neutralino annihilation, 
we use the fragmentation functions extracted from JETSET 7.4 \cite{LUND}.  
Since most of the $\bar{p}$'s are produced as the fragments of $b\bar{b}$ 
(no $\bar{p}$ from $\tautau$), the shape of their source spectrum does not 
depend much on the parameter set, although its absolute value scales with 
$S\Bbbbar$, where $S$ is the neutralino annihilation rate per unit volume 
and $\Bbbbar$ is the branching ratio into $b\bar{b}$.  The local 
interstellar flux of these $\bar{p}$'s is then calculated using a 3-D Monte 
Carlo simulation based on the diffusion model, as in the case of evaporating 
PBHs.  Note that, whereas the emission rate from evaporating PBHs per unit 
volume is proportional to their density, $S$ is proportional to the 
{\em squared} neutralino density; thus the local flux of $\bar{p}$'s from the 
neutralino annihilation is rather sensitive to the halo density distribution.  
  
We will show later that the flux of $\bar{p}$'s from the neutralino 
annihilation is usually too small to be observed under conventional 
astrophysical assumptions, i.e., if the neutralino has the homogeneous, 
spherical and isothermal distribution with $\rhalos=0.3$ GeV~cm$^{-3}$.  
There are, however, at least three astrophysical uncertainties which 
possibly increase the $\bar{p}$ flux by a factor of $\xi\gg 1$:  First, 
the Galactic halo may be flattened toward the Galactic disk \cite{R95}.  
This shall increase the local halo density $\rhalos$ by a factor of $\sim2$ 
over that of a spherical halo \cite{GGT95b}, resulting in the enhanced 
annihilation rate by a factor of $\xi\sim4$.  Secondly, a non-dissipative 
gravitational singularity (NGS) may reside at the Galactic center (GC) 
\cite{BGZ92}, resulting in the halo density distribution being 
$\rho_{\rm h}(r)=\rhalos(r/r_{\odot})^{-1.8}$ for $r\gtrsim0.1$ pc.  If 
this is the case, the neutralino annihilation rate would be very high at 
the GC\@.  Figure~\ref{fig1} shows the $r$-distribution of the primary 
sources (the neutralino annihilation and evaporating PBHs) which contribute 
to the integrated $\bar{p}$ flux at the Solar system.  It can be seen that, 
whereas only the annihilation occurring within a few kpc of the Solar 
system would contribute in the case of the isothermal density distribution, 
in the NGS model, $\bar{p}$'s from the annihilation at the GC would dominate 
the local interstellar $\bar{p}$ flux, leading to $\xi\sim200$.  Thirdly, 
high-density CDM clumps with a density enhancement factor of 
$10^{2}$--$10^{9}$ might be generated in the early Universe \cite{SS93}.  
If a small portion (a few \%) of neutralino dark matter is in the form of 
such clumps, the expected $\bar{p}$ flux could be enhanced by a factor of 
$\xi\gtrsim20$ \cite{BSS90}.  Furthermore, if there exists such a clump in 
the vicinity of the Solar system, the resultant $\bar{p}$ flux at the Earth 
could be further enhanced by a large factor.  

\section{Secondary \pbar ~flux}

When high-energy cosmic rays interact with the interstellar gas, secondary 
particles are produced including \pbar's.  The expected flux of such 
secondary \pbar's was calculated by a number of authors 
\cite{ste_pb,webb_pb,gaiss,simon_lb} who showed that the expected energy 
spectrum of the secondary \pbar's drops off steeply below 2 GeV. This 
feature originates from the kinematics of the \pbar ~production, and is 
independent of the details of the cosmic-ray propagation.

Recently, one of us has calculated the secondary \pbar ~fluxes \cite{Mitsui} 
using the two models of the cosmic-ray propagation, i.e., the Standard Leaky 
Box (SLB) \cite{eng} and Diffusive Reacceleration (DR) \cite{hein} models, 
by taking the most plausible input data such as the parent proton flux
\footnote[1]{The interstellar proton flux is taken to be
          $1.5 \times 10^{4} \beta ^{-1} P^{-2.74} \;
          {\rm m^{-2} sr^{-1} s^{-1} GeV^{-1}}$,
          where $P$ is the momentum in units of GeV,
          based on the recent observation of the
          LEAP experiment \cite{seoleap}.
         }.
The resultant spectra shown in Fig.~\ref{fig2} have very similar shapes, 
although the absolute flux in the SLB model is 2.5 times larger than that 
in the DR model.

\section{Solar modulation}

When antiprotons go into the Solar system, their energy spectra are changed 
due to the diffusion, convection, and deceleration processes by the solar 
wind and the interplanetary magnetic field (solar modulation).  The minimal 
model to describe the processes is the spherically symmetric model 
\cite{glee1,glee2}, in which all quantities are dependent only on the 
Heliocentric distance $\varrho$.  At the boundary of the Heliosphere 
($\varrho=\varrho_b$) the cosmic-ray energy spectrum is assumed to match the 
interstellar spectrum.  The energy spectrum at the Earth 
($\varrho=\varrho_E=1\;{\rm AU}$) can then be calculated by solving the 
diffusion-convection equation \cite{glee2} using the numerical technique 
developed by Fisk \cite{fisk}.

As usual, we take the solar wind speed $V=400$ km/s and the radius of the 
Heliosphere $\varrho_b = 60$ AU.  Since it was shown \cite{Mitsui} that the 
spectrum at the Earth is not significantly affected by the position 
dependence of the diffusion coefficient $\kappa$, we take $\kappa$ to be 
position independent and, as normally done, to be proportional to the 
particle rigidity $R$ as well as to the speed $\beta$, i.e., 
$\kappa = \beta \kappa_1 R$, where the coefficient $\kappa_1$ varies with 
solar activity.  The solar activity level is then represented by a single 
parameter:
$
\phi_F \equiv (\varrho_b - \varrho_E)V/3 \kappa_1,
$
which roughly corresponds to the ``$\phi$ parameter'' in the force field 
approximation \cite{glee2} 
\footnote[2]{
          Analyses of proton data show that $\phi_F \simeq \phi + 50$ MV.
         },
i.e., the energy loss in the Heliosphere per particle charge.  With $\phi_F$ 
varying between $\sim 350$ and $\sim 1500$ MV from the solar minimum to the 
maximum, this model reproduces well \cite{evens,seoleap,Mitsui} the large 
variation (a factor of 20) of low-energy proton and helium fluxes as well as 
each energy spectrum at various stages of solar activity.  Since the soft 
energy spectra of the primary $\bar{p}$'s are similar to those of proton and 
helium, we use this model of solar modulation to calculate the 
solar-modulated spectra of the primary $\bar{p}$'s.  

To forecast the time variation of $\phi_F$, we can utilize the data of the 
CLIMAX neutron monitor \cite{climax}, because the $\phi_F$ parameter shows 
the excellent correlation with the count rate of the neutron monitor 
\cite{evens}.  The CLIMAX data over the past 45 years clearly show 22-yr 
solar cycle.  Especially the recent trend of the data shows strong 
similarity to those in $\sim 1950$.  From this, one would expect the next 
solar minimum in the years 1997 to 1998 with the $\phi_F$ parameter reaching 
$\sim 350$ MV, which should be followed by a rapid increase of solar activity 
($\phi_F \sim 1000$ MV) in 2000.

\section{Results and discussion}\label{results}

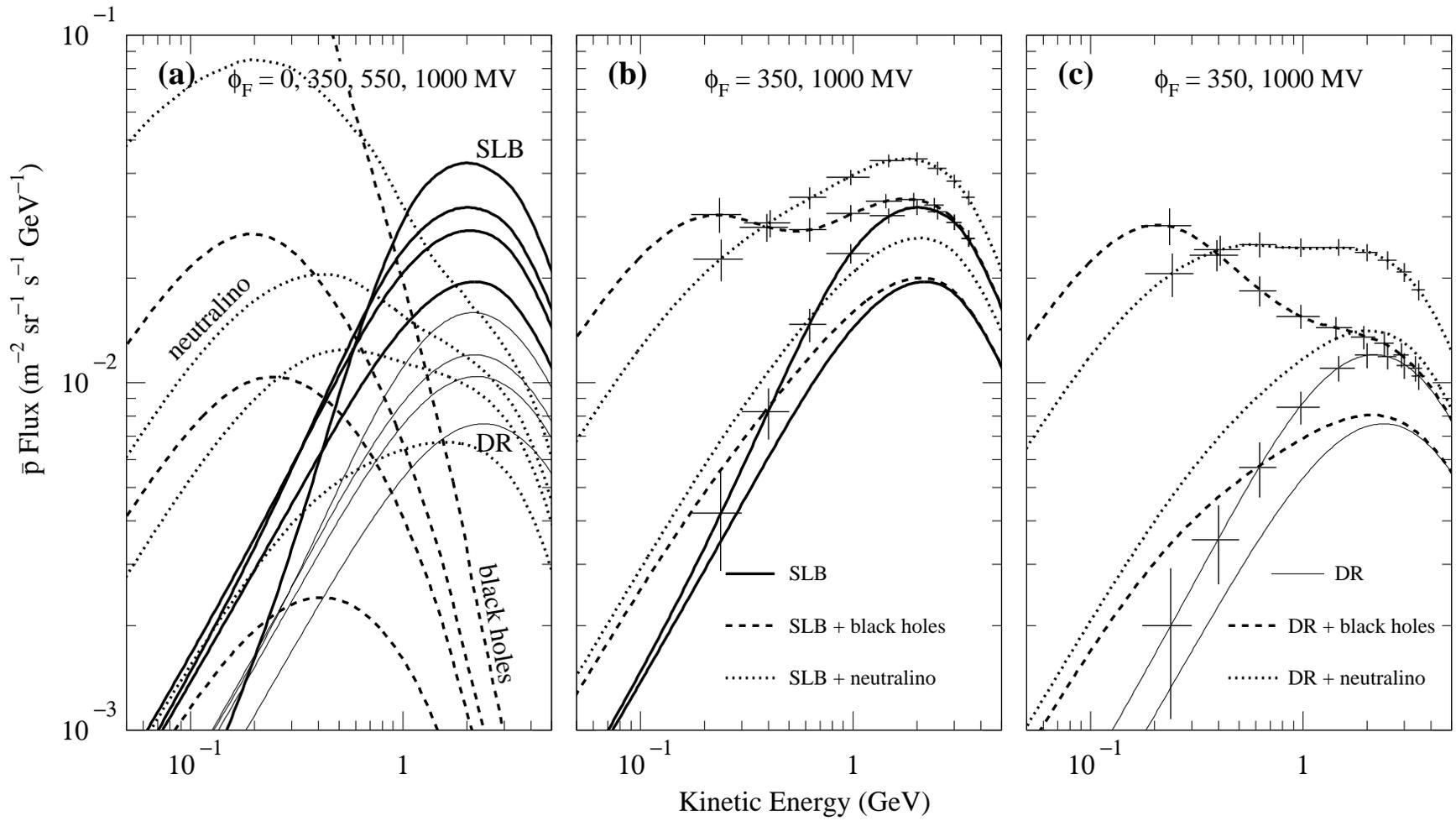
\begin{figure}[h]
\caption{
\protect\baselineskip 10mm
(a)
The expected energy spectra
of $\bar{p}$'s from evaporating PBHs
with ${\cal R}=5\times10^{-3}$ pc$^{-3}$yr$^{-1}$ (dashed lines)
and from the neutralino annihilation for the case~\#1 with $\xi=25$
(dotted lines), as well as the
secondary $\bar{p}$'s in the SLB
(thick solid lines) and DR (thin solid lines) models.
The curves correspond, from top to bottom, to
$\phi_F$ values of 0 (interstellar), 350, 550 and 1000 MV.
(b)
The expected spectra for the secondary \pbar ~(SLB) only (solid lines),
secondary \pbar ~(SLB) plus \pbar's from evaporating PBHs (dashed lines),
and secondary \pbar ~(SLB) plus \pbar's from neutralino annihilation
(dotted lines). The upper and lower curves correspond to
$\phi_F = 350$ and 1000 MV respectively.
The normalization parameters for the primary 
sources are the same as in (a).
Also shown is the expected statistical accuracy of a future
observation \cite{BESSlong}.
(c)
Same as (b) with the DR model of the secondary \pbar's.
}
\label{fig2}
\end{figure}

Figure \ref{fig2} (a) shows examples of the primary \pbar ~fluxes as well as 
the two secondary \pbar ~fluxes (SLB and DR) solar-modulated with 
$\phi_F=350$, 550, and 1000 MV, which would correspond to 1997, 1995, and 
2000 respectively.  The uppermost curves ($\phi_F = 0$) represent the 
interstellar spectra.  We have taken ${\cal R}=5\times10^{-3}$ 
pc$^{-3}$yr$^{-1}$ for the \pbar ~flux from evaporating PBHs, and the 
case~\#1 with $\xi=25$ for the \pbar ~flux from the neutralino annihilation.  
It should be noted \cite{imaxt} that the secondary \pbar ~fluxes are rather 
insensitive to the solar modulation. This is because the interstellar fluxes 
decrease with decreasing energy below 2 GeV so that the two effects of the 
solar modulation, i.e., deceleration and flux suppression cancel each other.  
Contrastingly, the primary \pbar ~fluxes would show a large enhancement 
toward the solar minimum in a very similar way to the proton flux.  Note that 
even the ``enhanced'' solar minimum fluxes of the primary \pbar's are very 
much suppressed as compared to the interstellar fluxes.

Based on the present understanding of the cosmic-ray propagation, we expect 
that the secondary antiprotons will most likely exist.  Therefore, we show in 
Fig.~\ref{fig2} (b) the solar-modulated ($\phi_F = 350$ and 1000 MV) 
\pbar ~fluxes in three cases, i.e., (i) secondary (in the SLB model) only, 
(ii) secondary (SLB) plus \pbar's from evaporating PBHs, and 
(iii) secondary (SLB) plus \pbar's from the neutralino annihilation.  
Figure \ref{fig2} (c) shows the corresponding spectra with the DR model of 
the secondary \pbar's.

The unknown absolute intensities of the primary \pbar ~fluxes (${\cal R}$ and 
$\xi$ for PBHs and neutralino dark matter respectively) were chosen here not 
to conflict with the existing observations.  Note that the value of 
${\cal R}$ taken here (see above) is 1/3 of the upper limit obtained in our 
previous paper \cite{MMO95} using the BESS '93 data \cite{BESS93}.  Such a 
weak signal, however, will appear as a striking enhancement of the 
low-energy \pbar ~flux at the solar minimum.  Also shown is the expected 
statistical accuracy estimated for a future measurement by the BESS 
experiment with a total 48-hour flight \cite{BESSlong}.  As seen in these 
figures, the existence of novel primary \pbar ~sources at this level could be 
detected without any ambiguity.  It can also be seen that the signals of the 
primary \pbar's will rapidly diminish after the solar minimum with the 
increase of solar activity.  This strong variation with solar activity 
would provide further evidence for the primary \pbar's, because the spectra 
of secondary \pbar's are not largely affected by the solar modulation.

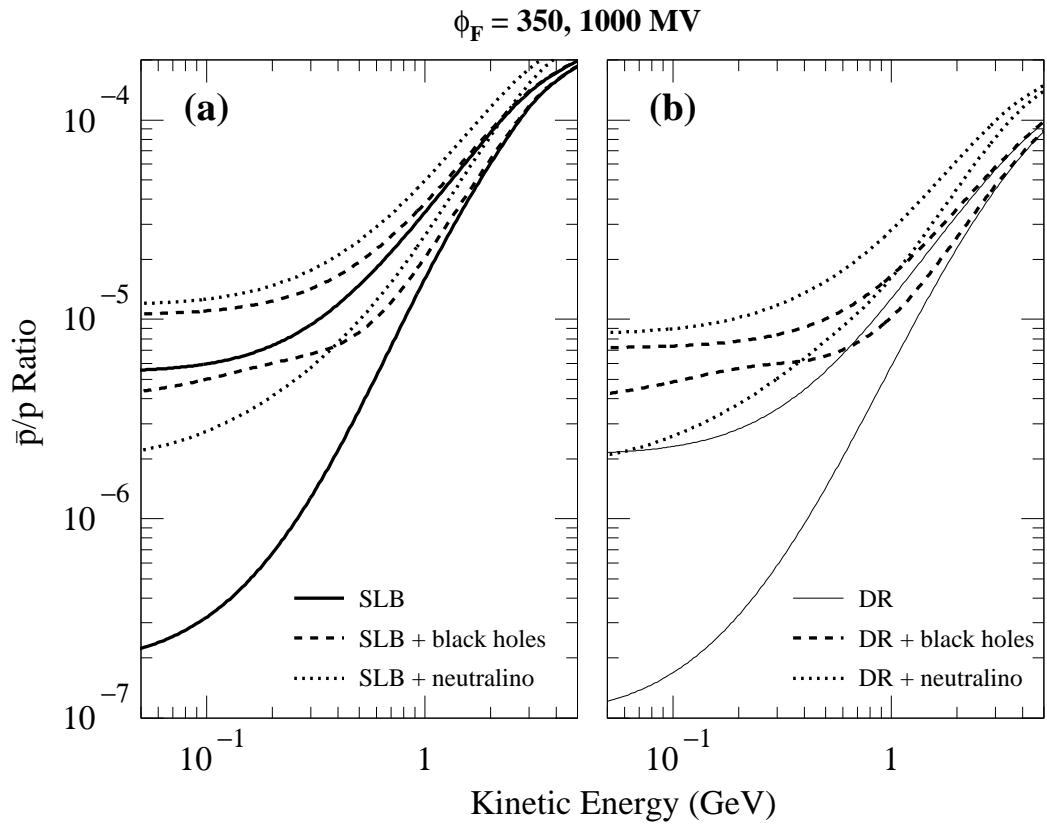
\begin{figure}[h]
\caption{
\protect\baselineskip 10mm
The $\bar{p}/p$ ratio corresponding to Fig.~\protect\ref{fig2}
(b) and (c). The lower and upper curves correspond to
$\phi_F = 350$ and 1000 MV respectively.
The proton spectra are obtained by solar-modulating
the interstellar spectrum:
$1.5 \times 10^{4} \beta ^{-1} P^{-2.74} \;
      {\rm m^{-2} sr^{-1} s^{-1} GeV^{-1}}$ \cite{seoleap},
where $P$ is the momentum in units of GeV.
}
\label{fig3}
\end{figure}

In the past, the observational data as well as theoretical calculations were 
usually presented in terms of the \pbar/$p$ ratio.  However, as shown in 
Fig.~\ref{fig3}, the $\bar{p}/p$ ratio is rather insensitive to the solar 
modulation of the primary \pbar's, since the denominator of proton flux 
varies in a similar way to the primary \pbar ~components.  Therefore, one has 
to measure the absolute \pbar ~flux and its spectrum in order to fully 
utilize the characteristic solar modulation of the primary \pbar's.

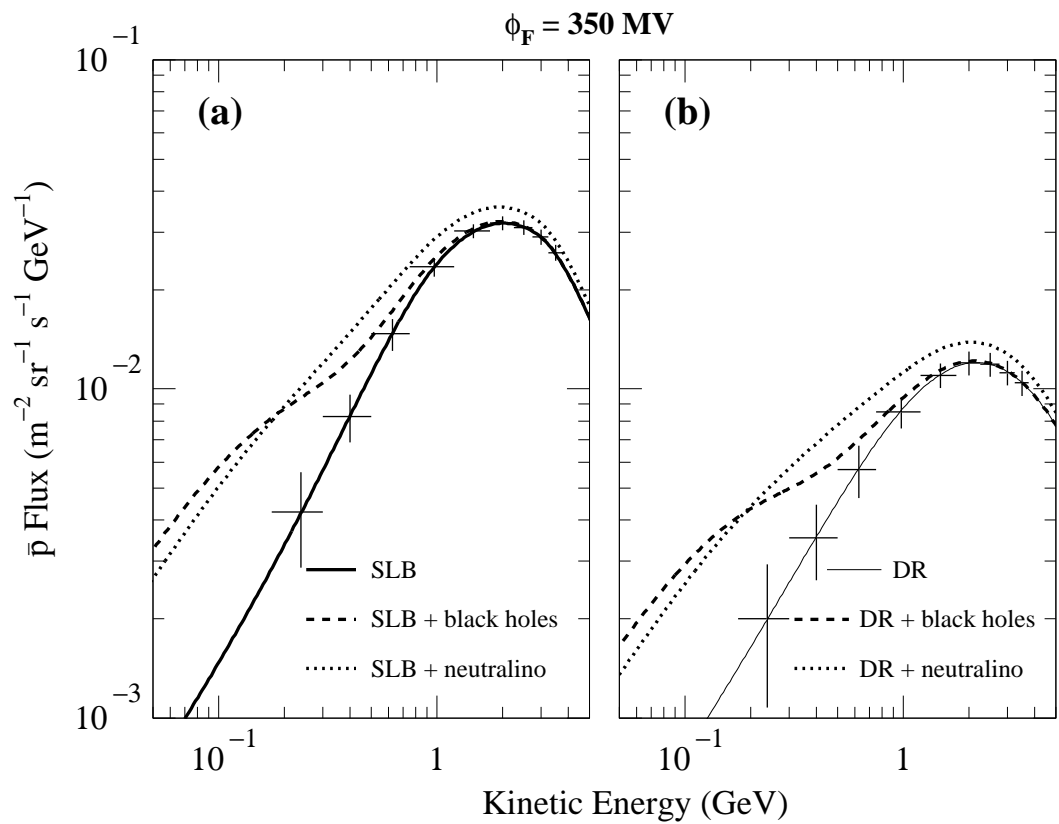
\begin{figure}[h]
\caption{
\protect\baselineskip 10mm
  (a)
  Same as Fig.~\ref{fig2} (b) but for
  ${\cal R}=1\times10^{-3}$ pc$^{-3}$yr$^{-1}$ and $\xi=8$.
  (b)
  Same as Fig.~\ref{fig2} (c) but for
  ${\cal R}=5\times10^{-4}$ pc$^{-3}$yr$^{-1}$ and $\xi=4$.
}
\label{fig4}
\end{figure}

Finally we discuss the attainable sensitivity to the primary 
\pbar ~components expected in the future experiments.  Figure \ref{fig4} (a) 
and (b) show the \pbar ~spectra corresponding to Fig.~\ref{fig2} (b) and (c), 
respectively, for much smaller contributions of the primary \pbar's.  As 
seen in the figures, these small signals can also be detected with the 
expected accuracy.  The corresponding sensitivity to the PBH explosion rate 
${\cal R}$ is $1\times10^{-3}$ pc$^{-3}$yr$^{-1}$ in case of the SLB model, 
and $5\times10^{-4}$ pc$^{-3}$yr$^{-1}$ in case of the DR model.  These 
values are respectively 1/17 and 1/34 of the best existing upper limit derived 
from the BESS '93 data.  Note that the explosion rate 
${\cal R}=1\times10^{-3}$ pc$^{-3}$yr$^{-1}$ is 9 orders of magnitude lower 
than the upper limit obtained from searches for TeV $\gamma$-ray bursts 
\cite{CYGNUS}, and is a few hundreds times lower than the upper limit derived 
from the anisotropy of the diffuse $\gamma$-ray flux \cite{W96}.  On the 
standard assumption of the PBH initial mass spectrum \cite{MMO95}, this value 
of ${\cal R}$ corresponds to the average PBH density in the Universe of 
$\Omega_{\rm PBH}=3\times10^{-10}$, which is a factor of $\sim30$ below the 
upper limit deduced from the cosmological diffuse $\gamma$-ray background 
\cite{PH76}.  

Figure~\ref{fig4} also shows that, in order for the $\bar{p}$ flux from the 
neutralino annihilation to be observable for the case~\#1, the enhancement 
factor $\xi$ must be greater than 8 and 4 when the SLB and DR models, 
respectively, are used for the secondary $\bar{p}$'s.  Since such enhancement 
is possible within astrophysical uncertainties as discussed in Section 3, 
$\bar{p}$'s from the neutralino annihilation could be observed in the near 
future.  

In conclusion, if future experiments allow us to precisely measure the 
low-energy cosmic-ray \pbar ~flux at the solar minimum, we should be able to
conduct a very sensitive search for
primary $\bar{p}$ components by utilizing the characteristic
solar modulation of the primary \pbar's.

\section*{Acknowledgements}
We are grateful to all the members collaborating on the BESS experiment.  
Sincere thanks are due to M. Drees and M.M. Nojiri for allowing us to use 
their routines to calculate the neutralino relic density.  The data of the 
CLIMAX neutron monitor were provided by University of Chicago, National 
Science Foundation Grant ATM-9420790.  All the calculations were performed 
on the RS/6000 workstations of the International Center for Elementary 
Particle Physics (ICEPP).  K.M. acknowledges a fellowship from the Japan 
Society for the Promotion of Science.  

\newpage

\newpage
\section*{Figure captions}

\newpage

\newpage
\section*{Tables}
\begin{table}[h]
\renewcommand{\arraystretch}{1.5}
\caption{
\protect\baselineskip 10mm
Representative sets of SUSY soft breaking parameters (assumed $A_{0}=0$) 
and relevant quantities derived from them, i.e., (i)~the neutralino mass, 
$\mchi$; (ii)~the mass of the lightest charged SUSY particle (LCSP), 
$\mlcsp$, together with its identity (chargino $\charg$ or stau $\stau$); 
(iii)~the neutralino annihilation 
rate per unit volume in the local region of the Galactic halo, 
$S\equiv(\rhalos/\mchi)^{2}\sigv$, where $\sigv$ is the thermally-averaged 
cross section for the neutralino annihilation in the Galactic halo; and 
(iv)~the branching ratio of the neutralino annihilation into bottom quarks, 
$\Bbbbar$.  Each set gives $\Ochi h^{2}=0.18$.  Masses $\mzero$, 
$\mhalf$, $\mchi$ and $\mlcsp$ are shown in units of GeV, $S$ in units of 
$10^{-32}$ cm$^{-3}$s$^{-1}$, and $\Bbbbar$ in units of \%.}
\label{tab1}
\vspace{8mm}
\begin{center}
\begin{tabular}{r|ccc|cccc}
\hline 
    & $\mzero$ & $\mhalf$ & $\tbeta$ & $\mchi$ & $\mlcsp$ & $S$ & $\Bbbbar$ \\
\hline
\#1 & 450 & 135 & 40 &  53.6 & 102 ($\charg$) & 32.7  & 96.6 \\
\#2 & 191 & 140 & 20 &  54.4 & 101 ($\charg$) & 18.9  & 97.6 \\
\#3 & 104 & 224 & 10 &  89.2 & 131 ($\stau$)  & 0.965 & 43.2 \\
\#4 & 107 & 165 &  5 &  62.7 & 117 ($\charg$) & 0.861 & 97.4 \\
\hline
\end{tabular}
\end{center}
\end{table}

\end{document}